\documentclass[letterpaper]{article} 
\usepackage{aaai2026}  
\usepackage{times}  
\usepackage{helvet}  
\usepackage{courier}  
\usepackage[hyphens]{url}  
\usepackage{graphicx} 
\usepackage{natbib}  
\usepackage{caption} 
\usepackage{algorithm}
\usepackage{algorithmic}

\usepackage{newfloat}
\usepackage{listings}
\DeclareCaptionStyle{ruled}{labelfont=normalfont,labelsep=colon,strut=off} 
\floatstyle{ruled}
\newfloat{listing}{tb}{lst}{}
\floatname{listing}{Listing}
 \nocopyright

\title{Neural Co-state Projection Regulator: A Model-free Paradigm for Real-time Optimal Control with Input Constraints}
\author{
    Lihan Lian\textsuperscript{\rm 1},
    Uduak Inyang-Udoh\textsuperscript{\rm 1}
}
\affiliations{
    \textsuperscript{\rm 1}University of Michigan\\

}

\usepackage{bibentry}

\usepackage{amsmath}
\usepackage{amssymb}  
\usepackage{booktabs}
\usepackage{array}

\begin{document}

\maketitle

\begin{abstract}

Learning-based approaches, notably Reinforcement Learning (RL), have shown promise for solving optimal control tasks without explicit system models. However, these approaches are often sample-inefficient, sensitive to reward design and hyperparameters, and prone to poor generalization, especially under input constraints. To address these challenges, we introduce the \textit{ neural co-state projection regulator} (NCPR), a \textit{ model-free} learning-based optimal control framework that is grounded in Pontryagin's Minimum Principle (PMP) and capable of solving quadratic regulator problems in nonlinear control-affine systems with input constraints. In this framework, a neural network (NN) is trained in a self-supervised setting to take the current state of the system as input and predict a finite-horizon trajectory of projected co‐states (i.e., the co-state weighted by the system's input gain). Subsequently, only the first element of the NN's prediction is extracted to solve a lightweight quadratic program (QP). This workflow is executed in a feedback control setting, allowing real-time computation of control actions that satisfy both input constraints and first‐order optimality conditions. 

We test the proposed learning-based model-free quadratic regulator on (1) a unicycle model robot reference tracking problem and (2) a pendulum swing-up task. For comparison, reinforcement learning is used on both tasks; and for context, a model-based controller is used in the unicycle model example. Our method demonstrates superior generalizability in terms of both unseen system states and varying input constraints, and also shows improved sampling efficiency.

\end{abstract}


\section{Introduction}
\paragraph{Motivation}
Feedback optimal control is useful for real-time decision-making in many critical systems as it enables autonomous adaptation of control actions to evolving state information in a manner that optimizes the system's performance while enforcing constraints \cite{teo2021applied}. However, nonlinear optimal control problems (OCPs) generally lack analytic solutions and have been classically solved numerically. Numeric methods are typically computationally intensive, as they require solving a new OCP at each feedback update \cite{peaucelle2010complexity}. To circumvent this computational burden, learning-based approaches, primarily reinforcement learning (RL), have emerged as alternatives \cite{sutton1998reinforcement}. However, while RL has the capacity to excel in learning optimal control policies even in black-box environments, it typically suffers from high sample inefficiency and lacks formal guarantees for stability or robustness to unseen conditions \cite{mediratta_generalization_2024, Henderson2018}.

\paragraph{Related Work} Numerical methods for solving OCPs fall into two categories: direct and indirect \cite{Betts2010}. Direct methods are typically implemented using model predictive control (MPC), in which the OCP is formulated as a nonlinear program that is solved in a receding horizon fashion \cite{Grune2011}. However, the nonlinear program may not converge rapidly enough to ensure feasibility for real-time feedback control \cite{schwenzer2021mpc}. Explicit MPC, which precomputes control laws offline, is only tractable for low-dimensional linear systems due to exponential memory and complexity requirements \cite{Nambisan2024Optimal}.  On the other hand, the indirect method derives the necessary conditions for optimality based on Pontryagin's Minimum Principle (PMP), resulting in a two-point boundary value problem (TPBVP) \cite{kirk2004pontryagin}. However, indirect methods are not suitable for feedback control as TPBVPs are often difficult to solve online, especially in the presence of state and control constraints \cite{rao2009survey,pagone2022penalty,pereira2021aggregated}. 

In addition to the computational drawbacks discussed above, numerical methods also require an accurate model of the system’s dynamics. In contrast, learning-based approaches, particularly reinforcement learning (RL), are capable of solving OCPs in a model-free fashion by leveraging dynamic programming principles to learn feedback policies directly through reward-driven interactions with the environment \cite{sutton1998reinforcement}. This model-free nature allows RL to excel in black-box or high-uncertainty settings, where numerical methods are difficult to apply. Several algorithms, such as Deep Deterministic Policy Gradient (DDPG) and Proximal Policy Optimization (PPO), have been widely used in many challenging control tasks \cite{ddpg, ppo}, including scenarios with system constraints \cite{Bhatia_Varakantham_Kumar_2019, Cheng_Orosz_Murray_Burdick_2019}. However, it remains difficult for a trained RL agent to transfer their experience to new environments, or generalize between tasks \cite{cobbe2019quantifying}

Recent studies have aimed to combine RL with model-based direct methods such as MPC by using a neural network (NN) to improve the design of cost functions \cite{rl_mpc_tnnls}. However, these approaches still require solving the receding-horizon OCP recursively, and thus suffer from the same computational burden as MPC. Another paradigm uses RL to learn low-level feedback control policies, identify a low-dimensional model of the resulting closed-loop system, and then generate a high-level model-based controller \cite{li2022bridging}. This approach, however, still suffers from aforementioned sample complexity associated with RL.

Other studies have sought to combine the indirect method with NN learning paradigms. Efforts in this direction have used NNs to approximate the TPBVP arising from PMP for specific initial conditions\cite{pontryagin_nn-mathematics, zang2022machine}. More recent work has introduced co-state neural networks (CoNNs), which parameterize the mapping from any given state to its corresponding optimal co-state trajectory using the supervision of expert TPBVP solvers \cite{lian2025costateneuralnetworkrealtime}. However, expert TPBVP solutions are generally suboptimal and non-unique, especially for higher-dimensional systems, making this supervised learning approach restrictive. Hence, a more recent work has proposed training the CoNN in a self-supervised manner such that the optimal solution minimizes a Hamiltonian function \cite{lian2025neuralcostateregulatordatadriven}. Nevertheless, both approaches rely on knowledge of the system's model. 


\paragraph{Contribution}
 In this work, we present a \textit{model-free} optimal control framework, \textit{neural co-state projection regulator} (NCPR), in which its core component, CPNN (Co-state Projection Neural Network), is trained in a self-supervised fashion, without an explicit system model. Based on the PMP, the CPNN is trained to find the mapping from a state to its corresponding optimal projected co-state trajectory. The control input can be subsequently obtained by solving a QP using the first entry of the CPNN prediction. To enumerate, the contributions of this paper is three-fold:
\begin{enumerate}
    \item We propose a novel \textit{model-free optimal control paradigm} using a PMP-informed loss function to train a CPNN. This eliminates the need for both exact system state space equations and solving the TPBVPs.

    \item We demonstrate the proposed NCPR framework has the ability to generalize to unseen initial states and nonzero references when compared with RL. The NCPR is also shown to have greater sampling efficiency and better flexibility in terms of handling input constraints. 
    

     \item We ascertain that the NCPR framework achieves performance comparable to that of the MPC with significantly reduced computational time.

\end{enumerate}

\section{Background}
Consider the finite-horizon, continuous time OCP:
\begin{subequations} \label{eq:background_ocp_formulation} 
\begin{align}
\mathcal{J} & = \phi(\mathbf{z}(t_f)) + \int_{0}^{t_f} L(\mathbf{z}(t), \mathbf{u}(t), t) \, dt, \\
\text{s.t.} \quad  & \dot{\mathbf{z}}(t) = f(\mathbf{z}(t), \mathbf{u}(t), t),  \label{eq:background_ocp_formulation-state_dynamics} \\
& \mathbf{z}(0) = \mathbf{z}_0, \label{eq:background_ocp_formulation-initial-condition}\\
& \mathbf{u}(t) \in \mathcal{U} \label{eq:background_ocp_formulation-input-constraint},
\end{align}
\end{subequations}
Here, $\mathbf{z}(t) \in \mathbb{R}^p$ denotes the state of the system (or tracking error), $\mathbf{u}(t) \in \mathbb{R}^q$ denotes the control input. The total cost $\mathcal{J}$ comprises a terminal cost term \(\phi(\mathbf{z}(t_f))\) and a stage cost integral \(L(\mathbf{z}(t), \mathbf{u}(t), t)\), evaluated over $[0,t_f]$. Any admissible state-control pair $\bigl(\mathbf{z}(\cdot),\mathbf{u}(\cdot)\bigr)$ must satisfy the constraints including system dynamics (\ref{eq:background_ocp_formulation-state_dynamics}), initial condition (\ref{eq:background_ocp_formulation-initial-condition}), and the input constraint (\ref{eq:background_ocp_formulation-input-constraint}).

\subsection{Solving OCP Numerically} 
Continuous time OCPs are typically solved by one of two broad classes of methods: Direct methods start by transcribing the OCP through time discretization. This is followed by solving an optimization problem with a finite number of decision variables, thus following the so-called \textit{discretize then optimize} paradigm. Indirect methods solve OCPs by first deriving the necessary conditions. This aligns with the paradigm of \textit{ optimize and then discretize} and preserves the analytic structure of the problem \cite{nonlinear-programming-book}. 

\subsubsection{Direct Method} 
For solving the OCP~\eqref{eq:background_ocp_formulation}, the direct method first discretizes the dynamics of the system as:
\begin{equation}
    \mathbf{z}_{k+1} = f(\mathbf{z}_k, \mathbf{u}_k),
\end{equation}
where $\mathbf{z}_k \in \mathbb{R}^p$ is the state vector, $\mathbf{u}_k \in \mathcal{U} \subset \mathbb{R}^q$ is the admissible control input, both at some time step $k = 0,\dots,N$ where $N$ is the total number of time steps. The cost is similarly discretized as 
\begin{equation}\label{discritized-cost}
\min_{\{\mathbf{z}_k, \mathbf{u}_k\}_{\forall k}} \quad \sum_{k=0}^{N-1} \ell(\mathbf{z}_k, \mathbf{u}_k)  + \phi(\mathbf{z}_N).
\end{equation}
Collocation or shooting methods are typically used to enforce system dynamics at each $k$, resulting in a nonlinear program (NLP) which may be solved using gradient-based algorithms\cite{rao2009survey}. In a feedback control setting, a new finite‑horizon NLP is solved at each time step \cite{Grune2011}. On obtaining the optimal control input sequence, only the first element is applied.

\subsubsection{Indirect Method} 
PMP is the foundation for this class of methods which provides first-order necessary conditions for optimality. Applying PMP to the OCP in ~\eqref{eq:background_ocp_formulation} introduces the control Hamiltonian $H$, defined as:
\begin{align}
H(\mathbf{z}(t), \mathbf{u}(t), \mathbf{\lambda}(t), t) &= L(\mathbf{z}(t), \mathbf{u}(t), t)\notag \\ 
 & + \mathbf{\lambda}^\top(t) f(\mathbf{z}(t), \mathbf{u}(t), t),
\label{eq:hamiltonian}
\end{align}
where \(\mathbf{\lambda}(t) \in \mathbb{R}^n\) denotes the trajectory of the co-state. The optimal control input $\mathbf{u}^*(t)$ can be obtained by solving an optimization problem while satisfying the following constraints on both the state and the co-state \cite{rao2009survey}:

\begin{equation}
\dot{\mathbf{z}}(t) = \nabla_{\mathbf{\lambda}} H,
\label{eq:state_dynamics}
\end{equation}
\begin{equation}
\dot{\mathbf{\lambda}}(t) =  -\nabla_{\mathbf{z}} H.
\label{eq:costate_dynamics}
\end{equation}
\begin{equation}
 \mathbf{\lambda}(t_f) = \nabla_\mathbf{z} \phi(\mathbf{z}(t_f)).
\label{eq:co-state_terminal_condition}
\end{equation}
\begin{equation}
\mathbf{u}^*(t) = \arg\min_{\mathbf{u}(t) \in \mathcal{U}} H(\mathbf{z}^*(t), \mathbf{u}(t), \lambda^*(t), t).
\label{eq:optimal_control}
\end{equation}
The OCP \eqref{eq:background_ocp_formulation} is of fixed final time but free final state, and the resulting TPBVP should then be solved by imposing the initial time boundary condition for the state and the final time boundary condition for the co‑state as shown in the Eq. \eqref{eq:co-state_terminal_condition}. When input constraints exist, the PMP states that optimal control \(\mathbf{u}^*(t)\) minimizes the Hamiltonian $H$ as indicated in Eq. \eqref{eq:optimal_control}. Closed‑form solutions to this TPBVP are generally unavailable. Numerical techniques, such as shooting methods, are commonly used \cite{numerical-tpbvp-book, collocation-method-ocp}.

\subsection{Reinforcement Learning}
Unlike direct and indirect methods, RL can tackle the same OCP in a model-free manner. Rather than relying on explicit knowledge of the dynamics \(f\), \(g\) and solving a TPBVP or nonlinear program, RL casts the problem as a Markov decision process (MDP) and learns optimal policies purely from sampled rollout based on the Bellman equation:
\begin{equation}\label{eq:bellman_equation}
V^*(s)
=\max_{a\in\mathcal{A}}
\bigl[r(s,a)
+\gamma\,\mathbb{E}_{s'\sim P(\cdot\mid s,a)}\bigl[V^*(s')\bigr]\bigr],
\end{equation}
where \(r(s,a)\) defines the reward, \(s\) is the current state, \(a\) is the chosen action, and \(s'\) is the successor state sampled from the probability distribution \(P(\cdot\mid s,a)\). This equation formalizes that the optimal value \(V^*(s)\) is equal to the best one-step reward plus the expected discounted value of the next state, where the discount factor $\gamma \in [0,1]$. RL algorithms then approximate either the state-value function $V^{\pi}(s)$ or action‑value function $Q^{\pi}(s,a)$:
\begin{equation}\label{eq:value_function}
    V^{\pi}(s)
=\mathbb{E}_{\pi}\Bigl[\sum_{k=0}^{\infty}\gamma^{k}\,r(s_{k},a_{k})\;\Big|\;s_{0}=s\Bigr],
\end{equation}
\begin{equation}\label{eq:q-function}
    Q^{\pi}(s,a)
=\mathbb{E}_{\pi}\Bigl[\sum_{k=0}^{\infty}\gamma^{k}\,r(s_{k},a_{k})\;\Big|\;s_{0}=s,\;a_{0}=a\Bigr],
\end{equation}

A canonical example is PPO \cite{ppo}, which alternates between collecting trajectories under current policy $\pi_{\theta}$. The learnable parameter $\theta$ is updated to maximize the clipped surrogate objective:
\begin{equation}\label{eq:ppo-loss-function}
L
=\mathbb{E}_{t}\Bigl[\min\bigl(r_{t}(\theta)\,A_{t},\,
\mathrm{clip}\bigl(r_{t}(\theta),1-\varepsilon,1+\varepsilon\bigr)\,A_{t}\bigr)\Bigr],
\end{equation}
where $A_{t}$ is an estimator of the advantage function, $
r_{t}(\theta)
=\frac{\pi_{\theta}(a_{t}\mid s_{t})}
      {\pi_{\theta_{\mathrm{old}}}(a_{t}\mid s_{t})}
$, and $\epsilon>0$ (typically $0.1$–$0.2$) defines the allowable deviation from the old policy. The clipping operation enforces $r_{t}(\theta)\in[\,1-\epsilon,\;1+\epsilon\,]$, approximating a trust‑region constraint that prevents overly large updates.  This mechanism enforces conservative policy updates that, to some extent, improve sample efficiency, making it a popular model-free algorithms for high‑dimensional control tasks.

\subsection{Bridging the Two Perspectives} 
The indirect method for solving OCPs aims to find a control input that satisfies the TPBVP as prescribed by the PMP. Meanwhile, the objective in RL is to find a control policy that maximizes a reward or minimizes the temporal difference (TD) loss without a system model.   In this work, we bridge both perspectives by introducing a framework in which the control agent aims to minimize the control Hamiltonian and satisfy the PMP. The proposed NCPR will replace the costly TPBVP solver with a CPNN that approximates the projected co‑state trajectory, enabling real‑time optimal control with satisfaction of the first-order necessary condition without knowledge of the system model.

\section{Problem Statement}
Consider a control-affine system for a finite-horizon OCP with a quadratic stage cost in continuous time. The objective is to minimize the cost functional: 
\begin{subequations} \label{eq:ocps_formulation} 
\begin{align}
\min \quad J & = \int_{0}^{t_f} \left( \mathbf{z^\top}Q\mathbf{z} + \mathbf{u^\top}R\mathbf{u} \right) dt + \phi(\mathbf{z}(t_f)), \\
\text{s.t.} \quad & \mathbf{\dot z}(t) = f(\mathbf{z}(t)) + g(\mathbf{z}(t))\mathbf{u}(t), \\
& \mathbf{z}(0) \in \mathcal{Z}\\ 
& \mathbf{u}(t) \in \mathcal{U}.  
\end{align}
\end{subequations}
Here, $\phi(\mathbf{z}(t_f))$ denotes the quadratic terminal cost by construction. The control input $u(t)$ is restricted to lie in the admissible set $\mathcal{U}$, while $\mathbf{z}(0)$ and $\mathbf{z}(t_f)$ denote the initial and terminal states, respectively. The set $\mathcal{Z}$ defines all allowable initial states\footnote{Note that the problem formulation here differs from standard OCPs, where $\mathbf{z}(0)$ is fixed. This is to emphasize that the OCP admits a family of solutions.}. The state space equations are described by the functions $f(\mathbf{z}(t))$ and $g(\mathbf{z}(t))$, with appropriate dimensions. The stage cost is quadratic, determined by the weighting matrices \(Q \in \mathbb{R}^{p \times p}\) and \(R \in \mathbb{R}^{q \times q}\), where $Q$ is a semi-definite symmetric matrix and $R$ is a symmetric positive definite matrix. 

Based on PMP, the Hamiltonian of OCP \eqref{eq:ocps_formulation} follows: 

\begin{align}
H(\mathbf{z}(t), \mathbf{u}(t), \mathbf{\lambda}(t), t) &= \mathbf{z}^\top(t)Q\mathbf{z}(t) + \mathbf{u}^\top(t)R\mathbf{u}(t) + \notag \\
& \mathbf{\lambda}^\top(t) \left( f(\mathbf{z}(t)) + g(\mathbf{z}(t))\mathbf{u}(t)\right),
\end{align}
where $\mathbf{\lambda}(t)\in \mathbb{R}^p$ is the co-state vector. From this Hamiltonian, the state and co-state equations follow: 
\begin{equation}
    \dot{\mathbf{z}}(t) = \nabla_{\mathbf{\lambda}} H = f(\mathbf{z}(t)) + g(\mathbf{z}(t))\mathbf{u}(t),
\end{equation}
\begin{align}
    \dot{\mathbf{\lambda}}(t) = -\nabla_{\mathbf{z}}{H} &= -2Q\mathbf{z}(t) - \nabla^\top_{\mathbf{z}}f(\mathbf{z}(t))  \mathbf{\lambda}(t) \notag \\
    &\quad -  \nabla^\top_{\mathbf{z}}\left(g(\mathbf{z}(t)) \mathbf{u}(t) \right) \mathbf{\lambda}(t).
\end{align}

After solving the resulting TPBVP for optimal $\mathbf{z}^*(t)$ and $\mathbf{\lambda}^*(t)$, one obtains the unconstrained optimal control law by enforcing
\begin{equation} \label{eq:unconstrained-optimal-u-partial-H-partial-u}
     \nabla_{\mathbf{u^*}}{H} = \mathbf{0} ,
\end{equation}
which yields
\begin{equation} \label{eq:unconstrained-optimal-u-expression}
    \mathbf{u}^*(t) = -\frac{1}{2} R^{-1}g^\top(\mathbf{z}(t)) \mathbf{\lambda}^*(t).
\end{equation}
When input constraints are active, one instead computes

\begin{equation} \label{eq:constrained-optimal-u}
\mathbf{u}^*(t) = \arg\min_{\mathbf{u}(t) \in \mathcal{U}} \left( \mathbf{u}^\top R\mathbf{u} + \mathbf{\lambda}^{*\top}(t) g(\mathbf{z}) \mathbf{u} \right).
\end{equation}

Note that for the TPBVP results from OCP \eqref{eq:ocps_formulation}, half of the boundary conditions are prescribed by the initial state, whereas the remaining conditions come from the co-state at the final time. To bypass the computational burden of numerically solving this TPBVP online, we introduce a CPNN that directly maps any admissible initial state $\mathbf{z}(0) \in \mathcal{Z}$ to its projected co-state trajectory $\mathbf{\lambda}^{*\top}(t) g(\mathbf{z})$, which suffices to obtain the optimal control input. 

\paragraph{Remark:} (1) Since the CPNN directly predicts $\mathbf{\lambda}^{*\top}_k g(\mathbf{z_k})$ at each time step $k$, and $R$ is defined in the cost function, this eliminates multiplication with the input gain matrix and the need for exact system dynamics (both $f$ and $g$).

(2) CPNN is trained in a self-supervised learning fashion, thus no expert TPBVPs sovler is needed.  Subsequently, this alleviates numerical inaccuracies arising from suboptimal TPBVP solutions, while ensuring that the NN generalizes across all $ \mathbf{z}(0) \in \mathcal{Z}$. 

\section{Methodology}
In this section, we first introduce the CPNN architecture and describe its training procedure. We then explain how the CPNN is used in a closed-loop feedback control setting to obtain control input that satisfies both constraints and first-order optimality, and how the algorithm is validated.

\subsection{Neural Network Architecture} 
The co-state projection neural network (CPNN) is implemented as a feedforward NN that maps an input state vector to a projected co‑state trajectory. At each time step $k$, for a state vector $\mathbf{z_k} \in \mathbb{R}^p$ and a finite horizon of CPNN prediction $n \in \mathbb{Z}^+$, the output of CPNN is the projection of $\boldsymbol{\hat{\lambda}^\top_{k}}$ on the input gain matrix $g(\mathbf{\hat{z}_k})$ for $i = k \dots k+n-1$,  

\begin{equation}
\boldsymbol{\hat{\Lambda}_{k}^\top} \circ g(\mathbf{\hat{Z}_k}) = \text{CPNN}_{\theta}(\mathbf{z_k}).
\end{equation}
Here, $\theta$ denotes the learnable parameters of the CPNN, and the operator $\circ$ is defined as follows:
\begin{equation}
    \boldsymbol{\hat{\Lambda}_{k}^\top} \circ g(\mathbf{\hat{Z}_k}) = 
    \begin{bmatrix}
      \boldsymbol{\hat{\lambda}^\top_{k}}g(\mathbf{\hat{z}_k})\\
    \boldsymbol{\hat{\lambda}^\top_{k+1}}g(\mathbf{\hat{z}_{k+1}})\\
      \vdots\\
      \boldsymbol{\hat{\lambda}^\top_{k+n-1}}g(\mathbf{\hat{z}_{k+n-1}})
    \end{bmatrix}.
\end{equation}
The co-state vector $\boldsymbol{\hat{\lambda}_{k}} = [\hat{\lambda}_{1,k}, \dots, \hat{\lambda}_{p,k}]^\top$ has a size of $p\times1$, and $g(\mathbf{\hat{z}_k})$ is of $p\times q$ for $i = k \dots k+n-1$. Thus, the dimension of the prediction of CPNN is $n\times q$. 

\subsection{Training Procedures} 
\begin{figure}[h!]
    \centering    
    \includegraphics[width=0.5\textwidth]{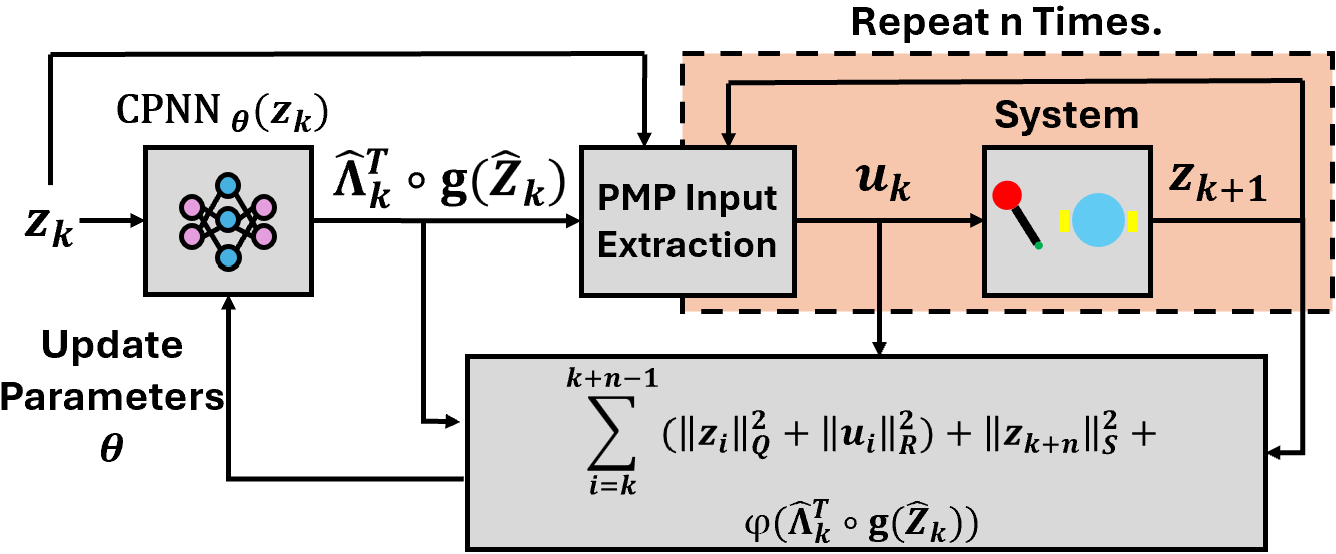}
    \caption{CPNN training pipeline.}
    \label{fig:training}
\end{figure}
\subsubsection{1) Training Data Generation} 
Let $\mathcal{Z}$ denote the admissible state space defined in OCP ~\eqref{eq:ocps_formulation}. We obtain a training set of $N$ 
data points by evenly spaced sampling over $\mathcal{Z}$. For the three-dimensional unicycle model used in the following example section, we choose $N = 1000 (10\times10\times10)$ sampled states for the training data set $\boldsymbol{z_{train}}$. For the two-dimensional pendulum system, we also generate $N = 100(10\times10)$ training data points in the same way.

\subsubsection{2) Loss function} 
Our training objective draws inspiration from the minimization of TD loss in Deep Q Learning \cite{deep-q-learning} but is tailored to similarly minimize the control Hamiltonian of the corresponding PMP formulation. We split the overall loss into three components:

\paragraph{1. Stage Loss} 
We accumulate the usual quadratic stage cost over the finite horizon:

    \begin{equation} \label{eq:L_stage}
    \mathcal{L}_{\text{stage}} = \sum_{i=k}^{k+n-1} \left( \mathbf{z_i}^\top Q\mathbf{z_i} + \mathbf{u_i}^\top R\mathbf{u_i} \right).
\end{equation}
Here, $\mathbf{u_i}$ can be obtained analytically based on Eq.~\eqref{eq:unconstrained-optimal-u-expression}. If CPNN is trained in a constrained manner, we simply clip $\mathbf{u_i}$ during training, as shown in Algorithm \ref{alg:cpn_training}.

\paragraph{2. Terminal Loss} 
 Similarly to the terminal cost in MPC, the second component is the terminal loss $\mathcal{L}_{\text{terminal}} = \phi(\mathbf{z_{k+n}})$. It is the same terminal cost function as defined in the corresponding OCP \eqref{eq:ocps_formulation}.

\paragraph{3. Regularization Loss} 
To drive the projected co-state trajectory to zero at the end of the prediction horizon, we consider two penalty designs of $\mathcal{L}_{\text{reg}} = \psi(\boldsymbol{\hat{\Lambda}_{k}^\top} \circ g(\mathbf{\hat{Z}_k}))$:
\begin{itemize}
    \item Uniform Penalty 
    \begin{equation*} \label{eq:L_co-state_simple}
\mathcal{L}_{\text{reg}} = \beta \|\boldsymbol{\hat{\Lambda}_{k}^\top} \circ g(\mathbf{\hat{Z}_k})\|_{1,1},
\end{equation*} 
which is defined as the sum of the absolute value of all projected co-state trajectory entries\footnote{For an $m\times n$ matrix $A$, we use the entry-wise matrix norm to define the loss as follows: $\|A\|_{p,p} = \|\text{vec}(A)\|_p = \left( \sum_{i=1}^{m} \sum_{j=1}^{n} |a_{ij}|^p \right)^{1/p}$.}, multiplied by a constant scalar $\beta$.

    \item Discounted Penalty
    \begin{equation*} \label{eq:L_co-state_complex}
\mathcal{L}_{\text{reg}} = \sum_{i=k}^{k+n-1} \left( \gamma^{\,k+n-i} \|\boldsymbol{\hat{\Lambda}_{k}^\top} \circ g(\mathbf{\hat{Z}_k})[i-k,:]\|_{1}\right),
\end{equation*}
where we reversely apply a discount factor $\gamma \in [0,1]$, which multiplied with the 1-norm of each entry of the CPNN prediction (projected co-state vector of size $1\times q$). This is to assign greater weight to the later entries of the projected co-state trajectory.
\end{itemize}

\begin{algorithm}[tb]
\caption{CPNN Training Procedures}
\label{alg:cpn_training}
\textbf{Input}: $\mathbf{z_k}$\\
\textbf{Parameter}: learning rate $\alpha$, training epochs $N_{epoch}$, CPNN prediction horizon $n$, learnable parameters $\theta$\\
\textbf{Output}: $\boldsymbol{\hat{\Lambda}_{k}^\top} \circ g(\mathbf{\hat{Z}_k})$
\begin{algorithmic}[1] 
\FOR{e in range($N_{epoch}$)}
    \FOR{$\mathbf{z_k}$ in $\boldsymbol{z_{train}}$}
    \STATE $\boldsymbol{\hat{\Lambda}_{k}^\top} \circ g(\mathbf{\hat{Z}_k}) = \text{CPNN}_{\theta}(\mathbf{z_k})$
    \STATE $\mathbf{\hat{U}} = [\mathbf{\hat{u}_{k}}, \dots \mathbf{\hat{u}_{k+n-1}}] = -\frac{1}{2} R^{-1}g^\top(\mathbf{\hat{Z}_k}) \boldsymbol{\hat{\Lambda}_{k}}$
    \IF {Constrained CPNN}
    \STATE $\mathbf{\hat{U}}$.clamp($u_{min}$, $u_{max}$)
    \ENDIF
    \STATE Calculate $\mathcal{L}_{\text{stage}}$, $\mathcal{L}_{\text{terminal}}$ and $\mathcal{L}_{\text{reg}}$
    \STATE \textbf{Update} CPNN parameters $\theta$
    \ENDFOR
\ENDFOR
\end{algorithmic}
\end{algorithm}

\subsection{Control Input Constraints Handling} 
Under the OCP formulation of ~\eqref{eq:ocps_formulation}, an unconstrained optimal control law at timestep $k$ follows Eq.~\eqref{eq:unconstrained-optimal-u-expression} directly, provided that the prediction of CPNN is optimal. When actuator limits are imposed, we instead solve a QP illustrated in Eq.~\eqref{eq:constrained-optimal-u}. Specifically, only the first element of the CPNN prediction is used, which is a $1\times p$ vector indexed by the $[0,:]$ operation. The QP is described as follows:
\begin{subequations} \label{eq:easier_qp} 
\begin{align}
    u^*_k &= \arg\min \left(\mathbf{u_k^\top} R \mathbf{u_k} + \boldsymbol{\hat{\lambda}_{k}^\top} g(\mathbf{\hat{z}_k}) \mathbf{u_k} \right), \\
    \text{s.t.} \quad & \boldsymbol{\hat{\Lambda}_{k}^\top} \circ g(\mathbf{\hat{Z}_k}) = \text{CPNN}_{\theta}(\mathbf{z_k}),  \\
    & \boldsymbol{\hat{\lambda}_{k}^\top} g(\mathbf{\hat{z}_k}) = \boldsymbol{\hat{\Lambda}_{k}^\top} \circ g(\mathbf{\hat{Z}_k})[0,:], \\
    & \mathbf{u_k} \in \mathcal{U}.
\end{align}
\end{subequations}

\subsection{Validation Pipeline} 
\begin{figure}[h!]
    \centering    
    \includegraphics[width=0.5\textwidth]{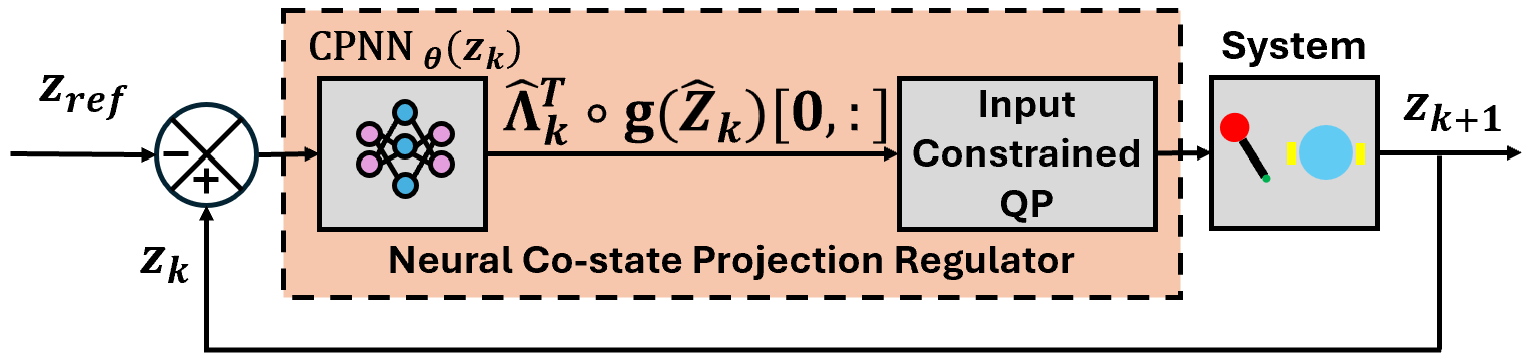}
    \caption{NCPR validation block diagram. Constrained optimal input is obtained by solving a QP that only use the first entry of CPNN prediction.}
    \label{fig:validation}
\end{figure}
Fig. \ref{fig:validation} depicts the integration of the CPNN into a validation pipeline. CPNN is evaluated in a real-time feedback control loop setting using simulation and at each time step $k$, the CPNN takes the state value $\mathbf{z}_k$ as input and predicts the corresponding optimal projected co-state trajectory $\boldsymbol{\hat{\Lambda}_{k}^\top} \circ g(\mathbf{z_k})$, which is a matrix of dimension $n\times q$. The QP based on Eq.~\eqref{eq:easier_qp} is then solved to obtain the control input. A fourth-order Runge-Kutta integrator then advance the system by one time step, and this whole process repeats continuously until the end.

\begin{figure*}[h!]
\centering
\includegraphics[width=0.98\textwidth]{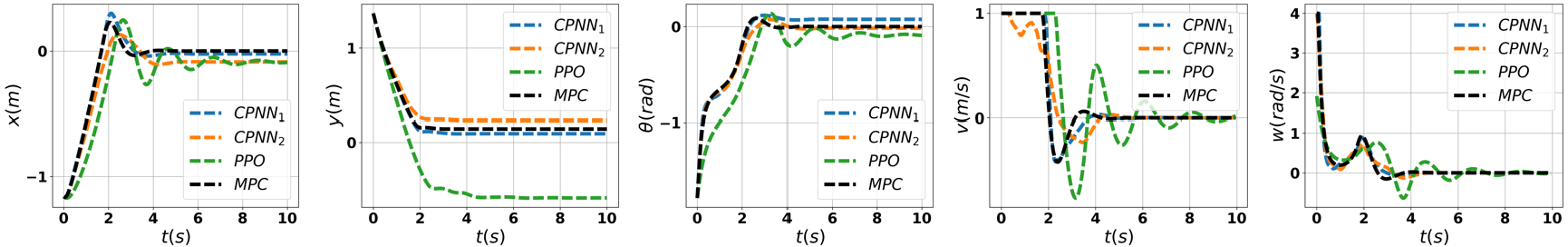} 
\caption{Comparison of solutions from four methods for $\mathbf{z}(0) = [-1.16, 1.37, -1.79]^\top$.}
\label{fig:unicycle_comparison_a}
\end{figure*}
\section{Examples}
We tested the performance of NCPR on two systems, specifically assessing its behavior under unseen initial states and nonzero references, as well as the effect of different input constraints during training and testing. All computations are completed on a computer with an i7 CPU and RTX 4070 GPU. $N_{epoch}=50$ for CPNN training in both examples , and the learning rate is $10^{-3}$ and $10^{-4}$ for unicycle and pendulum example respectively. For RL, PPO from Stable-Baseline3 \cite{stable-baselines3} is used for both examples.

\subsection{Unicycle Model}
Consider the following nonlinear OCP for a unicycle model, with the objective of minimizing a cost function below: 
\begin{subequations} \label{eq:unicycle_ocp} 
\begin{align}
\min_{u} \quad J  & = \int_{0}^{t_f} \left( \mathbf{z}^\top Q\mathbf{z} + \mathbf{u}^\top R\mathbf{u}\right) \, dt + \phi(\mathbf{z}(t_f)),\\
\text{s.t.} \quad &  \dot{z_1} = \dot{x} = v \cos(\theta),\\
&  \dot{z_2} = \dot{y} = v \sin(\theta),\\
&  \dot{z_3} =\dot{\theta} = \omega,\\
& \mathbf{u} \in \mathcal{U}, \\
& \mathbf{z}(0) \in \mathbb{R}^3. 
\end{align}
\end{subequations}

Here, $\mathbf{z}=[x, y, \theta]^\top$ denotes the state of the system, and the control input $\mathbf{u} = [v, w]^\top$ consists of the linear velocity $v$ and the angular velocity $w$. The control inputs are restricted to $-1 \leq v \leq 1,  -4 \leq \omega \leq 4,$ in accordance with the setting in \cite{rl_mpc_tnnls}. We set $Q = diag(10,10,10), R = diag(1,1)$ for the stage cost term and for the terminal cost $\phi(\mathbf{z}(t_f)) = \mathbf{z}^\top(t_f)S\mathbf{z}(t_f)$, we choose $S = 50Q=diag(500,500,500)$. 

In the simulation set-up, we use the sampling time $dt = 0.05 s$ and prediction horizon $n = 30$ for both CPNN and MPC. Thus, $t_f = 1.5s$ in this case. \textit{CasADi} is employed as the optimization framework for MPC, with \textit{ipopt} selected as the NLP solver. During the CPNN training stage, we choose the loss function with discounted penalty for the regularization loss $\mathcal{L}_{\text{reg}}$ and $\gamma = 0.99$. A total of 1000 ($10 \times 10 \times10$) states are used as training data, and 10 samples are evenly spaced from $[-2, 2]$ for all $x, y$ and $\theta$. A total of $1.5\times10^6(30\cdot1000\cdot50)$ time steps are used to train both CPNN ($CPNN_1$) and constrained CPNN ($CPNN_2$). 

For RL training, the reward functions are designed as the negative of the stage cost in the OCP \eqref{eq:unicycle_ocp}, with a horizon $t_f=10s$. This setup is designed to approximate the infinite-horizon OCP, and thus no terminal reward is employed. The same number of training time steps is used for PPO.

\paragraph{Seen Initial State}
The performance of NCPR is first evaluated under an initial condition that lies within the training data distribution, specifically $\mathbf{z}(0) = [-1.16, 1.37, -1.79]^\top$. As shown in Fig. \ref{fig:unicycle_comparison_a}, a total of four methods are compared. PPO exhibits the worst performance in terms of convergence error, which is defined as the sum of absolute tracking error across all state variables, as indicated in Table \ref{unicycle_comparison_table}. The $CPNN_1$ achieves superior tracking performance compared to the constrained $CPNN_2$. Both CPNNs show a result comparable to MPC, but with significantly lower computational cost, with a speed of $1.6 ms$ per simulation step, in contrast to $241.1 ms$ for MPC.   

\begin{figure*}[t]
\centering
\includegraphics[width=0.98\textwidth]{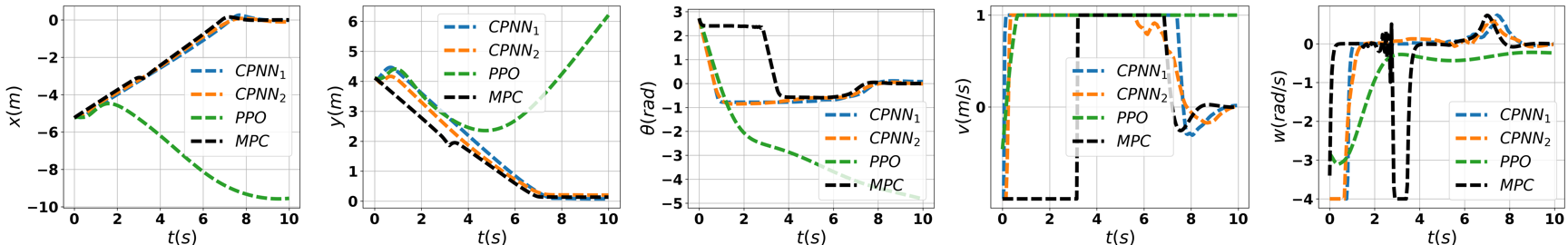} 
\caption{Comparison of solutions from four methods for OOD $\mathbf{z}(0) = [-5.24, 4.11, 2.72]^\top$.}
\label{fig:unicycle_comparison_b}
\end{figure*}

\paragraph{Unseen Initial State}
To validate the generalizability of CPNN, we further assess performance under an out-of-distribution (OOD) initial condition, $\mathbf{z}(0) = [-5.24, 4.11, 2.72]^\top$. As shown in Fig. \ref{fig:unicycle_comparison_b}, PPO fails to guide the robot to the target zero reference state. Although MPC has the smallest convergence error, it does so at a substantially higher computational cost of $282.3ms/step$, compared to $1.6 ms/step$ for both CPNN methods. Moreover, MPC generates control input trajectory that exhibit the least trajectory smoothness, as measured by mean squared derivatives (MSD)\footnote{The mean squared derivative is computed by first calculating the numerical gradient at each point using the \textit{np.gradient} function, squaring each value, summing them, and dividing by the total number of points.}, due to abrupt changes shown in Fig. \ref{fig:unicycle_comparison_b}.

\begin{figure*}[h!]
\centering
\includegraphics[width=0.98\textwidth]{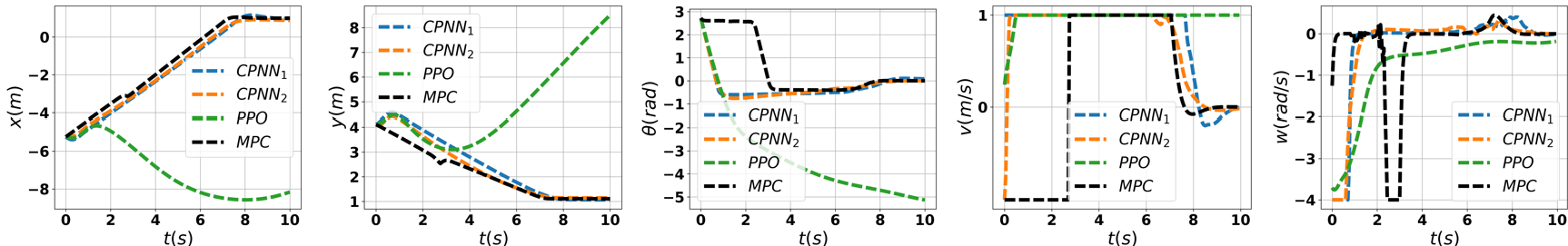} 
\caption{Comparison of solutions from four methods for OOD $\mathbf{z}(0) = [-5.24, 4.11, 2.72]^\top$ with nonzero reference $[1, 1, 0]^\top$.}
\label{fig:unicycle_comparison_c}
\end{figure*}

\begin{table}[h!]
\centering
\begin{tabular}{ c c c c}    
    \toprule
     & \textbf{Case A} & \textbf{Case B} &\textbf{ Case C} \\
    \midrule
    \shortstack{$\textbf{CPNN}_1$ \\ (NCPR)} & \shortstack{Error: 0.19\\MSD: 2.92} & \shortstack{Error: 0.17\\MSD: 6.17} & \shortstack{Error: 0.17\\MSD: 3.73} \\
    \midrule
    \shortstack{$\textbf{CPNN}_2$ \\ (NCPR)} & \shortstack{Error: 0.33\\MSD: 1.75} & \shortstack{Error: 0.32\\MSD: 2.72} & \shortstack{Error: 0.26\\MSD: 2.53} \\
    \midrule
    \shortstack{\textbf{PPO} \\ (RL)} & \shortstack{Error: 0.78\\\textit{MSD: 0.83}} & \shortstack{Error: 20.58\\\textit{MSD: 0.40}} & \shortstack{Error: 21.8\\\textit{MSD: 0.34}} \\
    \midrule
    $\textbf{MPC}$ & \shortstack{\textit{Error: 0.14}\\MSD: 2.21} & \shortstack{\textit{Error: 0.14}\\MSD: 17.6} & \shortstack{\textit{Error: 0.11}\\MSD: 10.04} \\
    \bottomrule
\end{tabular}
\caption{Comparison table for unicycle model tasks. \textit{Italicize} entries indicate the best performance.}
\label{unicycle_comparison_table}
\end{table}

\paragraph{Unseen Initial State and Nonzero Reference}
We retain the same OOD initial state $\mathbf{z}(0) = [-5.24, 4.11, 2.72]^\top$, but assign the system with reaching a nonzero reference state $\mathbf{z_{ref}} = [1, 1, 0]^\top$. The input of the CPNN then becomes the error state ($\mathbf{z_k} - \mathbf{z_{ref}}$). PPO again fails to drive the robot to the target state, while CPNN still demonstrates a convergence error comparable to that of MPC, with smoother input trajectories. MPC has the best tracking performance, but with a computational cost of $292.4ms/step$, whereas NCPR consistently maintains a speed of $1.6ms/step$. All resulting state and control input trajectories are shown in Fig. \ref{fig:unicycle_comparison_c}.

\subsection{Pendulum}
The proposed algorithm is further tested on the pendulum swing-up control task, which can be described as follows:
\begin{subequations} \label{eq:pendulum_ocp} 
\begin{align}
\min_{u} \quad J  & = \int_{0}^{t_f} \left( \mathbf{z}^\top Q\mathbf{z} + \mathbf{u}^\top R\mathbf{u}\right) \, dt + \phi(\mathbf{z}(t_f)),\\
\text{s.t.} \quad &  \dot{z_1} = \dot{\theta},\\
&  \dot{z_2} = \ddot{\theta} = -\frac{gsin(\theta)}{l} + \frac{1}{ml^2} u,\\
& \mathbf{u} \in \mathcal{U}, \\
& \mathbf{z}(0) \in \mathbb{R}^2. 
\end{align}
\end{subequations}

Here, the dynamical system has two state variables $\theta$ and $\dot{\theta}$, and one control input $\tau$. Two sets of input constraints are considered: a wide constraints $-10 < \tau < 10$, and a tighter constraints $-2 < \tau < 2$. We choose $Q=diag(100,100)$, $R=1$ for the stage cost and for the terminal cost, $\phi(\mathbf{z}(t_f)) = \mathbf{z}^\top(t_f)S\mathbf{z}(t_f)$, with $S = 10Q$. We set $dt = 0.05s$, and for pendulum configuration, $m=1kg$, $l=1m$ and $g=9.81m/s^2$. CPNN training data are sampled from $[-2,2]$ for both $\theta$ and $\dot{\theta}$, with 10 evenly spaced data points for each state variable. Thus, a total of $100 (10\times10)$  state vectors are used. PPO is used for RL training and $\mathbf{z}(0)$ is uniformly sampled from the same range. The reward function is defined as the negative of the stage cost defined in \eqref{eq:pendulum_ocp}, and the duration of the episode is fixed to $t_f  = 10 s$, without any terminal reward. This setup allows the episodic reward to approximate the corresponding infinite-horizon OCP. 

 For both $CPNN_1$ and $CPNN_2$, the prediction horizon is set to $n=20$ and we used a uniform penalty for $L_{reg}$, with $\beta=0.1$. A total of $10^5(20\cdot100\cdot50)$ total time steps is used for both CPNN, while a total of $10^6$ time steps are used to ensure convergence for PPO. In the following examples, for $CPNN_2$, we use the tighter constraint $-2 < \tau < 2$ during training, while using the wider constraint $-10 < \tau < 10$ during testing. For PPO, the input constraint $-10 < \tau < 10$ is enforced throughout both training and testing.

\begin{figure}[H]
    \centering    
    \includegraphics[width=0.48\textwidth]{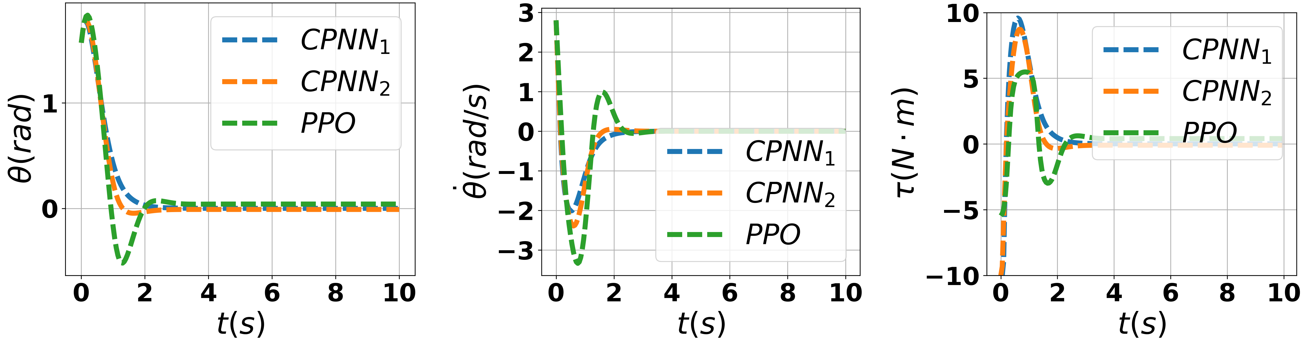}
    \caption{State and input trajectories comparison for pendulum swing-up task, where $\dot{\theta}$ is OOD ($\mathbf{z}(0)= [1.57, 2.8]^\top$).}
    \label{fig:pendulum_comparison_a}
\end{figure}

\begin{figure}[H]
    \centering    
    \includegraphics[width=0.48\textwidth]{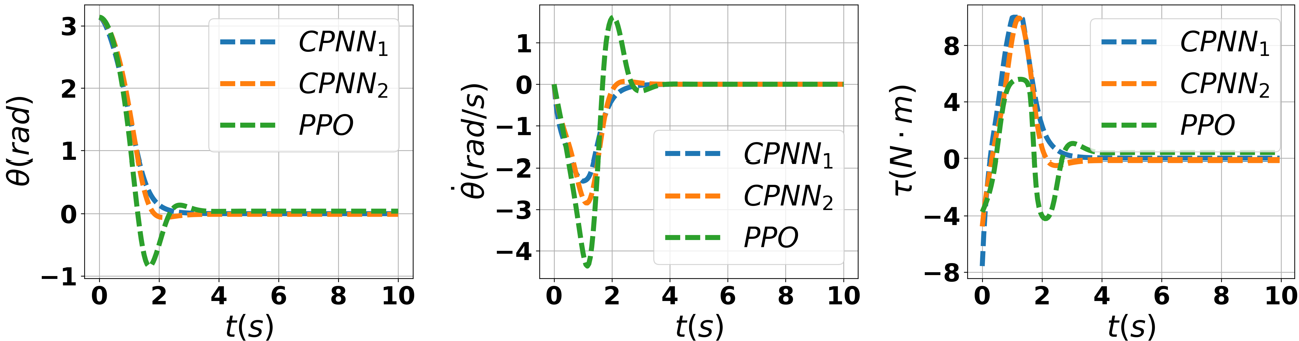}
    \caption{State and input trajectories comparison for pendulum swing-up task, where $\theta$ is OOD ($\mathbf{z}(0)= [3.14, 0.0]^\top$).}
    \label{fig:pendulum_comparison_b}
\end{figure}

\paragraph{One unseen initial state variable}
To assess the generalization capability of the controllers when only one state variable is OOD, we evaluate two scenarios. In the first case, $\dot{\theta}$ is OOD and $\mathbf{z}(0)=[1.57, 2.8]^\top$. In the second case, $\theta$ is OOD, with $\mathbf{z}(0)=[3.14, 0]^\top$. In both scenarios, $CPNN_1$ achieves zero convergence error, $CPNN_2$ results in the $0.01$ convergence error, and PPO produces the highest value of $0.04$ in both cases, as shown in Figs. \ref{fig:pendulum_comparison_a} and \ref{fig:pendulum_comparison_b}. 

Furthermore, although PPO uses consistent input constraints $[-10, 10]$ during both training and testing and has more training time steps, it has the least optimal result. Even $CPNN_2$, which is trained with the tighter constraints of $[-2,2]$, outperforms PPO. This is evidenced by PPO's pronounced overshoot in both $\theta$ and $\dot{\theta}$ trajectory, as well as inefficient use of control effort. Thus, PPO is not only less efficient in training, but also exhibits inferior generalizability under varying input constraints.

\begin{figure}[H]
    \centering    
    \includegraphics[width=0.48\textwidth]{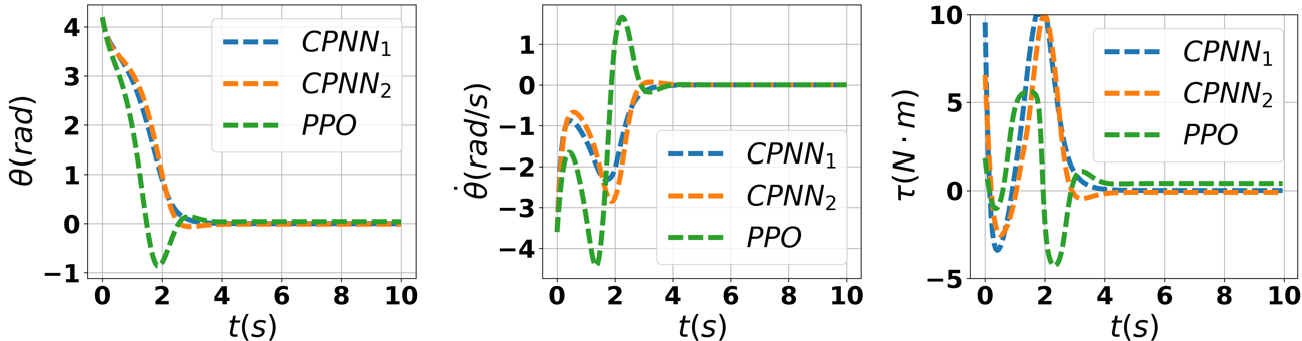}
    \caption{State and input trajectories comparison for pendulum swing-up task, where both $\theta$ and $\dot{\theta}$ are OOD ($\mathbf{z}(0)= [4.2, 3.6]^\top$).}
    \label{fig:pendulum_comparison_c}
\end{figure}

\paragraph{Two unseen initial state variable}
We also test the case where both $\theta$ and $\dot{\theta}$ are outside the range of $[-2, 2]$, with $\mathbf{z}(0)=[4.2, -3.6]^\top$. As illustrated in Fig. \ref{fig:pendulum_comparison_c}, all methods successfully regulate the pendulum to the target state. The convergence error for $CPNN_1$, $CPNN_2$ and PPO is $0, 0.01$ and $0.04$, respectively. PPO again has the largest overshoot in both $\theta$ and $\dot{\theta}$ trajectories, along with the lowest utilization of control effort, indicating the least optimal performance. Compared to $CPNN_1$, $CPNN_2$ achieves roughly the same performance, but with slightly larger overshoot, especially for the $\dot{\theta}$ trajectory.

\section{Conclusion}

Based on the foundation of PMP, we present a learning-based \textit{model-free} control algorithm, which does not require the known state-space equation like canonical optimal control techniques. The core component of our proposed \textit{neural co-state projection regulator} (NCPR), the co-state projection neural network (CPNN), is trained in a self-supervised manner and bypasses the need for a TPBVP solver. 

Our NCPR shows comparable results with MPC in the unicycle example, but with a much faster computational speed, and is more capable of handling input constraints in the pendulum example compared with RL. NCPR also demonstrates better generalization capability and greater sampling efficiency in both examples. Future work may be designing the algorithm for the predictive control task, as the current method is just a regulator. The better design of the NN architecture or the loss function could be another perspective of improvement.

\bibliography{aaai2026}

\end{document}